# Spintronics And Ferromagnetism In Wide-Band-Gap Semiconductors

Tomasz Dietl

*Institute of Physics, Polish Academy of Sciences and ERATO Semiconductor Spintronics Project of Japan Science and Technology Agency, al. Lotników 32/46, PL 02 668 Warszawa, Poland*

**Abstract.** Recent progress in understanding and controlling spintronic properties of (Ga,Mn)As and related compounds is contrasted to diverging experimental and theoretical results concerning the origin of high temperature ferromagnetism discovered in an ample class of magnetically doped wide-band-gap semiconductors.

## INTRODUCTION

In course of the years spintronics research has spread over all branches of condensed matter physics and materials science. Indeed, because of asymmetry in abundance of electric and magnetic elementary charges, random magnetic fields are weaker than random electric fields, so that the electron spin may at the end be better information carrier than the electron charge in both classical and quantum information technologies. With no doubt the most advanced are studies on metal magnetic multilayers, in which spin dependent electron scattering and tunnelling are employed in reading heads of hard disks and in random access magnetic memories (MRAMs) that are now reaching the production stage. However, particularly interesting appear ferromagnetic semiconductors as they combine resources of magnetic and semiconducting systems. Nevertheless, so far only displays of (Zn,Mn)S and Faraday optical insulators of (Cd,Mn)Te and related compounds are the working semiconductor spintronic devices. Now, research on semiconductor spintronics [1-4] is at this fascinating point, where many conceptual, fabrication, and characterization challenges are ahead of us.

In this paper, I summarize recent progress in understanding basic properties of (Ga,Mn)As and related materials [5-7], and then discuss issues we encounter developing wide-band-gap semiconductor systems in which spontaneous magnetization persists to above room temperature [8-11]. I do not describe advances in spintronics research on non-magnetic semiconductors [1-4], where strong spin-orbit interaction in narrow-gap semiconductors can serve for spin manipulation, while its weakness in wide-gap semiconductors, together with a small magnitude of dielectric constant, offer a worthwhile opportunity for developing functional quantum gates.

## CARRIER-CONTROLLED FERROMAGNETISM IN DMS

Owing to a short-range character of the direct exchange interaction between tightly localized magnetic orbitals, the coupling between d spins proceeds indirectly, *via* sp bands in tetrahedrally coordinated diluted magnetic semiconductors (DMS) [12-14]. Such a coupling is usually antiferromagnetic if the sp bands are either entirely occupied or totally empty (superexchange) but can acquire a ferromagnetic character in the presence of free carriers (Zener/RKKY mechanism). The discovery of carrier-induced ferromagnetism in zinc-blende III–V compounds containing a few percent of Mn [16,17] in which $T_C$ can exceed 100 K [18] followed by the prediction [19] and the observation [20,21] of ferromagnetism in *p*-type (II,Mn)–VI materials, opened up new areas for exploration. We can now consider physical phenomena and device concepts of previously unavailable combinations of quantum structures and magnetism in semiconductors [22]. In particular, since in these systems magnetic properties are controlled by the holes in the valence band, the powerful methods developed to change the carrier concentration by light and electric field in semiconductor structures can be employed to alter the magnetic ordering. Such tuning capabilities were demonstrated in (In,Mn)As/(Al,Ga)Sb [23-25] and (Cd,Mn)Te/(Cd,Zn,Mg)Te:N [20,26] heterostructures, one example is displayed in Fig. 1. Importantly, the magnetization switching is isothermal, reversible, and fast. Simultaneously, the injection of spin-polarized holes from (Ga,Mn)As to (In, Ga)As quantum wells in *p-i-n* light-emitting diodes (LEDs) was demonstrated [27,28]. The injection of spin-polarized electrons, using Zener or Esaki tunneling from *p*-type (Ga, Mn)As electrodes into *n*-type GaAs, was also realized [28-30]. At the same time, outstanding phenomena known from the earlier studies of metallic multilayer structures were observed in ferromagnetic DMSs, including switching of magnetization by current pulses [31,32], interlayer coupling [33,34], exchange bias-type behavior [35], giant magnetoresistance (GMR) [33], and tunneling magnetoresistance (TMR) [36,37]. It is, therefore, the important challenge of materials science to understand ferromagnetism in these compounds.

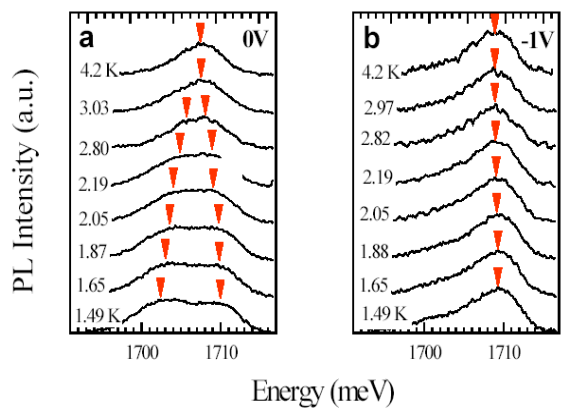

**FIGURE 1.** Photoluminescence at various temperatures of a p-i-n diode containing (Cd,Mn)Te quantum well sandwiched between *p*- and *n*-doped (Cd,Zn,Mg)Te barriers. In the absence of bias, the quantum well is modulation-doped by holes, which mediate ferromagnetic interaction between Mn spins, so that a spontaneous line splitting appears below 3 K (a). The ordering of Mn spins vanishes if the quantum well is depleted by the application of 1 V in the reverse direction (b) (after Boukari et al. [26]).

Figure 2 singles out Mn containing semiconductors, in which ferromagnetic coupling mediated by holes in the valence band has been put into evidence. In these materials, the Mn d-level resides deeply in the valence band [39,40] but owing to a large on-site Hubbard repulsion energy it remains half-filled and thus provides the localized spin $S = 5/2$. Since the d level is deep in the valence band, Mn behaves electrically like Zn -- it acts as an isoelectronic impurity in II-VI's but forms a fairly shallow acceptor or double acceptor state in III-V and group IV semiconductors in question, respectively. Similarly to other doped semiconductors, if an average distance between acceptors becomes about 2.5 times smaller than the corresponding Bohr radius, the Anderson-Mott insulator-to-metal transition occurs. In II-VI DMS [12-14] doping by N or P acceptors has been employed [19,20,40,41] to induce the ferromagnetic coupling, in some analogy to EuO [42] and (Pb,Mn,Sn)Te [43], where the carrier concentration, and thus ferromagnetism was controlled by annealing.

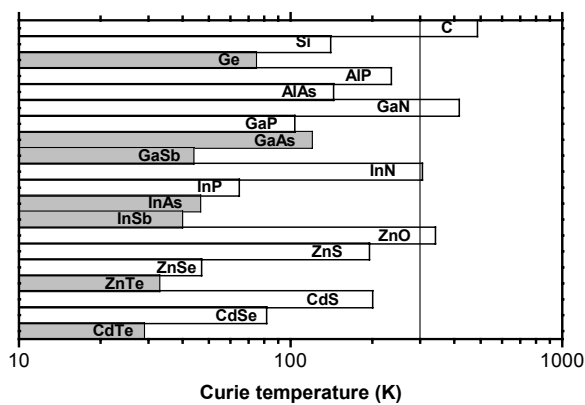

**FIGURE 2.** Diagram of Curie temperatures computed within the Zener model for various semiconductors assuming that 5% of cations are replaced by Mn and that the hole concentration is $3.5 \times 10^{20}$ cm$^{-3}$. The shadowed bars correspond to the materials for which the applicability of the model has been experimentally verified (adapted from Refs. [44,45]).

Taking the above experimental facts into account, theory Mn-based DMS has been built on Zener's model of ferromagnetism, the Ginzburg-Landau approach to phase transitions, the Kohn-Luttinger *kp* theory of semiconductors, and the Kubo formulation of charge transport phenomena [19,44-50]. It has been found that the Zener theory describes chemical trends in $T_C$, its scaling with the carrier and Mn spin concentrations as well as temperature and field dependent magnetization, hole spin polarization, strain-induced magnetic anisotropy, magnetic stiffness, stripe domain width, anisotropic magnetoresistance, anomalous Hall effect, magnetic circular dichroism, and a.c. conductivity of (Ga,Mn)As. Furthermore, effects of quantum localization has been assessed showing that the magnitude of the RKKY coupling behave in a non-critical way on crossing the metal-to-insulator transition [51] as well as demonstrating that the negative magnetoresistance at $T \ll T_C$ is caused by the *orbital* weak localization effect [52]. An interplay between the short-range antiferromagnetic coupling and long-range RKKY interactions in II-VI DMS has also been investigated [53]. It has also been found, in agreement with the theoretical expectations [19], that the electrons can lead to ferromagnetism only under rather extreme conditions [40,54].

## TOWARDS ROOM TEMPERATURE SEMICONDUCTOR SPINTRONICS

In view of potential for novel devices and system architectures, it has become timely to develop functional semiconductor systems with Curie temperatures $T_C$ comfortably exceeding room temperature, and in which semiconductor and magnetic properties can be controlled on an equal footing. Two strategies were proposed to accomplish this objective [44]. First was to increase the Mn and hole concentrations in the established carrier-controlled ferromagnets, such as (Ga,Mn)As.

Second was to develop new DMSs characterized by strong p-d hybridization, large density of states, and weak spin-orbit interactions, in which $T_C$ can be large even at moderate values of $x$ and $p$. It was, however, obvious that such a program would face a number of obstacles such as self-compensation, solubility limits, and tight binding of holes by magnetic ions in the limit of a strong p-d coupling [44].

Figure 3 presents state of the art results for (Ga,Mn)As with the peak value of $T_C$ =173 K at $x$ = 0.08 [55], slightly above earlier findings [56,57]. As shown, $T_C$ growths significantly upon low temperature annealing, an effect associated with the out-diffusion and surface oxidation of Mn interstitials [58]. This particular point defect was recently identified to account for self-compensation [59-62]. Its concentration increases, therefore, with the deepness of the Fermi level in the valence band during the epitaxy, that is with $p$ and $x$ [62]. Furthermore, according to the growth diagram depicted schematically in Fig. 4, a temperature window for epitaxy of $Ga_{1-x}Mn_{1-x}As$ shrinks with $x$, so that it is difficult to avoid the presence of roughening or MnAs precipitates when attempting to go beyond $x$ = 0.08. MnAs nanocrystals embedded in the GaAs matrix can be also obtained by appropriate post-growth annealing of (Ga,Mn)As at high temperatures [63,64]. As shown in Fig. 5, despite their small dimensions, the apparent $T_C$ of MnAs precipitates is close to the bulk value if they appear in a hexagonal form or is even higher if a zinc-blende structure is stabilized by the host [64]. Interestingly, the presence of precipitates is visible neither by HRXRD [64] nor in transport properties [5] but affects strongly magnetic circular dichroism [65].

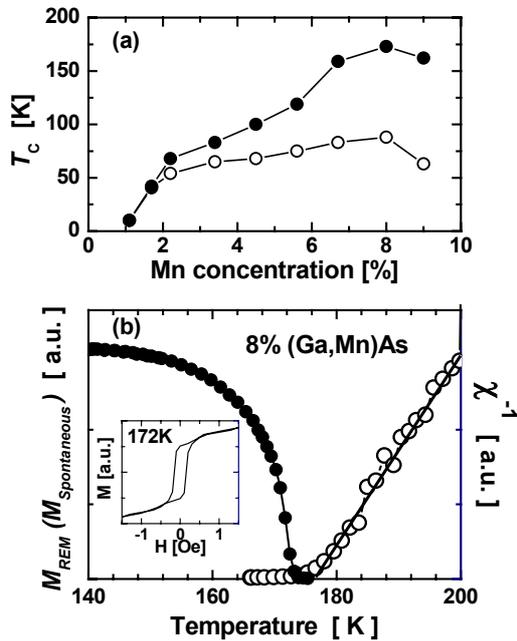

**FIGURE 3.** Curie temperatures for as-grown and annealed samples (empty and full, symbols, respectively) Remanent magnetization (full symbols), inverse magnetic susceptibility (empty symbols), and hystersis loop obtained from SQUID measurements for $Ga_{0.92}Mn_{0.08}As$ film after low temperature annealing. The data point to Curie temperature of 173 K, close to Curie-Weiss temperature (after Wang et al. [55]).

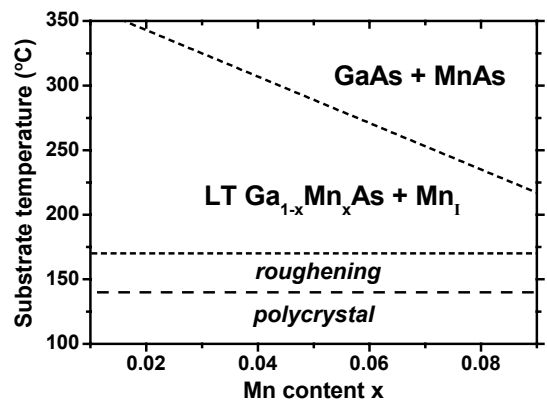

**FIGURE 4.** Schematic diagram of temperature window for growth of (Ga,Mn)As epilayers by low temperature (LT) molecular beam epitaxy. With the increase of the Mn content $x$ the window shrinks and the concentration of Mn interstitials increases (adapted from Matsukura et al. [5]).

The predicted high values of $T_C$ for p-type (Zn,Mn)O and (Mn,III)N, as shown in Fig. 2, stimulated a considerable growth effort. Interestingly, spontaneous magnetization persisting well above $T_C$ of $Ga_{1-x}Mn_xAs$, usually involving only a fraction of the spins, has been observed in a number of magnetically doped nitrides, oxides, and other wide-gap semiconductors [8-11], even without any holes in the valence band. Furthermore, a number of *ab initio* works has been carried out within the LSDA [66,67] substantiating the experimental observations. In most cases, the computed ferromagnetism has not been associated with the presence of the carriers in the sp bands but resulted from the double exchange – hopping of d electrons among the magnetic ions.

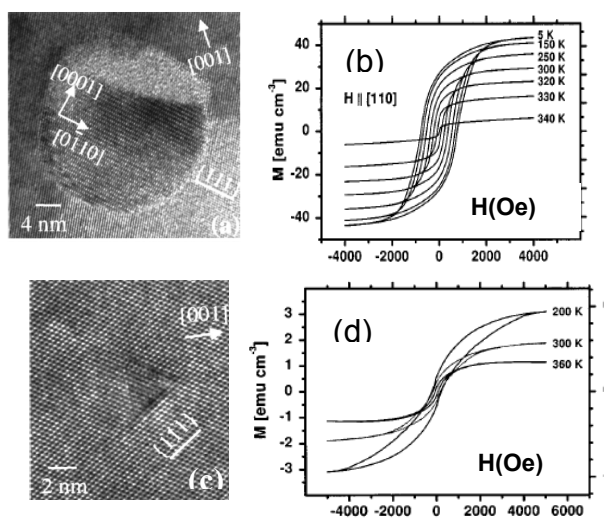

**FIGURE 5.** Transmission electron micrographs and hystersis loops showing persistence of spontaneous magnetization to above 300 K of hexagonal (a,b) and zinc-blende (c,d) MnAs nanoprecipitates in GaAs obtained by annealing of LT (Ga,Mn)As (after Moreno *et al.* [64]).

Unfortunately, however, the employed *ab initio* computation schemes overestimate tendency towards the ferromagnetic ground state by underestimating significantly the localization of the d electrons by the Mott-Hubbard and Anderson-Mott mechanisms, the latter often referred to as the magnetic-bond percolation problem. Actually, the presence of net ferromagnetic superexchange had been predicted for $d^3$ and $d^4$ ions in zinc-blende DMS employing a more reliable *empirical* tight-binding approach [68]. However, the expected magnitude and range of the interaction appear too small to describe high temperature ferromagnetism of DMS containing usually low concentration of magnetic constituent. How experimental facts can therefore be explained? I believe that there is no single model capable of interpreting the whole body of findings. However, just to show possible directions of reasoning, I will consider a few examples.

In the case of chalcopyrite $(Zn,Mn)GeP_2$, a ferromagnetic order appears above 47 K and persists up to about 300 K[69]. Surprisingly, this complex phase diagram is independent of the Mn concentration. We note, however, its striking similarity to that of MnP, where a screw antiferromagnetic phase develops below 47 K, while at higher temperatures a ferromagnetic phase with $T_C$ = 291 K is observed [70]. One can, therefore, conclude with a high degree of certainty that MnP precipitates control magnetic properties of $(Zn,Mn)GeP_2$.

Another relevant material is (Zn,Cr)Te, in which $T_C$ was found to increase with x and to attain 300 ± 10 K for x = 0.2 [71]. Unfortunately, however, in both (Zn,Cr)Te [72] and (Zn,Cr)Se [73], the Curie-Weiss temperature stays almost constant as a function of x and by no means shows the expected linear decrease with decreasing x, the dependence expected rigorously for a random distribution of magnetic ions. The body of experimental results suggests the presence of a superparamagnetism, presumably due to nanoclusters of ferromagnetic Cr spinels [74] or chalcogenides [75], whose size increases with x.

Despite investigations by many groups [8,9,76], the origin of high temperature ferromagnetic signal in (Ga,Mn)N is still unknown. Its presence does not seem to correlate with the Fermi level position. If, therefore, the relevant ferromagnetic coupling is not mediated by free carriers, its range should be rather small. Accordingly, no ferromagnetism is expected below the percolation limit, which is 19.5% for the nearest neighbors and 13.6% for the next nearest neighbors. Since the high temperature ferromagnetism is observed even if Mn content is of the order of 2%, once more a presence of precipitates, say of ferrimagnetic $Mn_4N$ [77-79], can be suggested.

Another interesting compound is (Zn,Co)O, in which a clear correlation between the amount of a ferromagnetic phase (or magnetic moment per Co) and oxygen pressure during laser ablation has been established [80,81]. On the other hand, n-type doping by Al appear to have a little effect. Since precipitates of Co or its compounds have not been detected so far, oxygen-vacancy- [81] or hydrogen-related defects [82] are being considered. It can be pointed out that vacancies may affect ferromagnetism indirectly, for instance, by enhancing atomic diffusion and, thus, facilitating aggregation of magnetic particles.

Finally, we mention (Zn,Fe,Cu)O, in which at a given Fe concentration, the magnitude of spontaneous magnetization increases with the Cu content [83]. An appealing possibility here is that because of the coincidence in energy of the Fe donor state with the Cu acceptor state [6] a channel for d electron hopping opens up, in other words, the two-ion double exchange operates [84].

## CONCLUSIONS

According to results described here, (Ga,Mn)As and related compounds emerge as the best understood ferromagnets. In contrast, we are at the beginning of the road to understanding and controlling magnetism of those wide-band-gap semiconductors, which support spontaneous magnetization up to high temperatures. Whether the observed effects are produced by precipitates of known ferromagnetic or ferrimagnetic materials or rather signalize novel physics, such as magnetism without magnetic elements [85], is to be elucidated in the years to come.

## ACKNOWLEDGMENTS

The work was partly supported by AMORE (GRD1-1999-10502) and FENIKS (G5RD-CT-2001- 00535) EC projects, by Humboldt Foundation, and carried out in collaboration with M. Sawicki, P. Kossacki, and T. Andrearczyk in Warsaw, as well as with groups of H. Ohno in Sendai, J. Cibert in Grenoble, B. Gallagher in Nottingham; and A.H. MacDonald in Austin.